\documentclass[10pt]{article}
\usepackage{dcolumn}% Align table columns on decimal point
\usepackage{bm}% bold math
\usepackage{verbatim}       % Defines \begin{comment}, \end{comment}

\usepackage[dvips]{graphicx}
\usepackage{amscd}
\usepackage{amsthm}
\usepackage{amssymb}

\usepackage{amstext}
\usepackage{psfrag}
\usepackage{amsmath}
\usepackage{amsfonts}
\usepackage{units}
\usepackage{mathrsfs}
\newcommand{\p}{\partial}

\newcommand{\dd}{{\rm d}}

\newcommand{\bd}{\begin{definition}}                %inizia definizione
\newcommand{\ed}{\end{definition}}                  %fine definizione
\newcommand{\bc}{\begin{corollary}}                 %inizia corollario
\newcommand{\ec}{\end{corollary}}                   %fine corollario
\newcommand{\bl}{\begin{lemma}}                     %inizia lemma
\newcommand{\el}{\end{lemma}}                       %fine lemma
\newcommand{\bp}{\begin{proposition}}            %inizia proposizione
\newcommand{\ep}{\end{proposition}}                %fine proposizione
\newcommand{\bere}{\begin{remark}}                  %inizia osservazione
\newcommand{\ere}{\end{remark}}                     %fine oservazione

\newcommand{\bt}{\begin{theorem}}
\newcommand{\et}{\end{theorem}}

\newcommand{\be}{\begin{equation}}
\newcommand{\ee}{\end{equation}}

\newcommand{\bit}{\begin{itemize}}
\newcommand{\eit}{\end{itemize}}
\newtheorem{theorem}{Theorem}[section]
\newtheorem{corollary}[theorem]{Corollary}
\newtheorem{lemma}[theorem]{Lemma}
\newtheorem{proposition}[theorem]{Proposition}
\theoremstyle{definition}
\newtheorem{definition}[theorem]{Definition}
\theoremstyle{remark}
\newtheorem{remark}[theorem]{Remark}
\newtheorem{example}[theorem]{Example}

\begin{document}
%
%\DeclareGraphicsExtensions{.pdf}

%\title{Connection between Lorentzian distance and mechanical least action in spacetimes admitting a parallel null vector}

\title{Inclusion of a perfect fluid term into the Einstein-Hilbert action}

\author{E. Minguzzi \footnote{Dipartimento di Matematica Applicata, Universit\`a degli Studi di Firenze,  Via
S. Marta 3,  I-50139 Firenze, Italy. E-mail:
ettore.minguzzi@unifi.it}}

%\pacs{04.20.Gz, 04.30.-w}

\date{}
\maketitle

\begin{abstract}
\noindent
I introduce a method  to obtain the stress-energy tensor of the perfect fluid
by adding a suitable term to the Einstein-Hilbert action. Variation should be understood with respect to the metric.
\end{abstract}

\section{Introduction}
Since the gravitational side of the Einstein equations is variational one would like to obtain the right-hand said from the variation of a matter Lagrangian.
This raises the problem as to whether the dynamical equations for the most straightforward form of matter, the perfect fluid, are variational in character. These equations, namely the continuity and the  Euler equations have been given variational formulation following different approaches \cite{taub54,hawking73,schutz77,baylin80,brown93,poplawski09,minazzoli12}.

The most common strategy  \cite{taub54,carter73,kijowski79,kunzle84,carter89,kijowski90,carter94,comer93,kijowski98,hajicek98,dunbosky06,slobodeanu11,alba15} considers  a submersion $\xi\colon M\to B$ to the  body frame. The fibers $\xi^{-1}(b)$  represent the flow lines. The action depends on the flow lines namely on the map $\xi$, i.e.\ the coordinate functions $\xi^A$, and on their spacetime derivatives  $\xi^A_\mu$.
In this fashion the dynamics of the continua admits a field theoretical formulation. Different type of continua are described by different geometric structures placed on $B$, for instance,  in a fluid $B$ would be endowed with a volume form $r(\xi) \dd \xi^1\cdots\wedge \dd \xi^n$ (telling us the amount of matter in a portion of the continua), while in an elastic material $B$ would be endowed with a metric $\gamma$ (telling us the distance between particles in their rest state). In all cases $B$ inherits a time dependent contravariant metric (telling us the distance between particles on spacetime) by push forward of the contravariant spacetime metric $G^{-1}=\xi_* g^{-1}$, i.e.\ $G^{AB}=g^{\mu \nu} \xi^A_\mu \xi^B_\mu$. The density can then be shown to be  $\rho= r \sqrt{\det G^{AB}}$, and natural Lagrangians can be constructed as functions of $G^{AB}$ or, in the case of a perfect fluid, of $\rho$, $L=F(\rho)$. In fact one can show that  variation with respect to the metric returns the stress-energy tensor of the perfect fluid.

This approach is natural but somewhat elaborated. It is necessary to introduce the projection $\xi$ and to write an action dependent on the derivatives of such projection, though one is really interested on dynamical equations which do not involve these variables. This drawback has motivated some authors to look for alternative approaches \cite{ootsuka16,ariki16}. %It also motivated me
 %which for presence of constraints, lack of elegance or other drawbacks had not convinced all authors. There is continuing activity on this topic as it is showed by an interesting work by Ootsuka at all \cite{}  recently posted on the archive.
%This interest prompted me to present
%I wish here to present this little note in which
 %This motivated me
% to present

 In this little note  I show that the perfect fluid stress-energy tensor can be obtained variationally in a more elementary and direct way. The derivation is really easy and has turned out to be essentially the same of Schutz and Schmid \cite{schutz70,schmid70}. Though at present this work is not meant for publication, it could still be useful as an introduction to the topic.
As with some other references \cite{carter73,comer93,slobodeanu11},  I take  the view that the variation should be taken with respect to the metric, and I shall not consider variation with respect to the flow lines.
%This will be a substantial difference with some previous works.

There are three reasons for this choice. Firstly this is the type of variation needed in the variational formulation of gravity coupled with matter. Secondly, while there is evidence that at the fundamental level matter is composed by particles described by vector fields which obey variational equations, there is no reason of principle to believe that {\em continua} should have a dynamical variational description. Indeed, the process of averaging needed to obtain the continua description might lead to 'averaged' equations which, though coming from variational equations, might not be themselves variational. Finally, it is known that variation with respect to $u$ and other thermodynamic quantities cannot give the correct dynamical equations without the introduction of constraints \cite{schutz77}. On the contrary, we shall not need to introduce neither constrains nor Lagrange multipliers.

\subsection{Completing the Einstein-Hilbert action}

%Let the spacetime dimension be $n+1$.
Let us denote for short  $\sqrt{-g} \,\dd^{4} x \to \dd x$, and let us adopt the conventions of \cite{misner73} (metric signature $(-,+,+,+)$, units chosen so that $c=G=1$). The Einstein equations are
\[
G_{\alpha \beta}+\Lambda g_{\mu \nu} =8 \pi T_{\alpha \beta} ,
\]
where if variational the stress-energy tensor $T_{\alpha \beta}$ is identified with
\[
T_{\alpha \beta}=g_{\alpha \beta}L_{(m)}-2 \frac{\p L_{(m)}}{\p g^{\alpha \beta}},
\]
where $L_{(m)}$ is the matter Lagrangian.
In this case  the variational principle is
\[
S=\int \left(\frac{1}{16\pi }(R-2 \Lambda) +L_{(m)}\right) \dd x ,
\]
indeed the variation gives
\begin{align*}
\delta S&=\int \left(-\frac{1}{16\pi } (G^{\mu \nu}+\Lambda g^{\mu \nu})+\frac{1}{2}L_{(m)} g^{\mu \nu}-\frac{\p L_{(m)}}{\p g^{\alpha \beta}}\, g^{\alpha \mu} g^{\beta \nu} \right) \delta g_{\mu \nu}\, \dd x,
%\\
%&=\int \left(-\frac{1}{16\pi } (G^{\mu \nu}+\Lambda g^{\mu \nu})+\frac{1}{2}L_{(m)} g^{\mu \nu}+\frac{\p L_{(m)}}{\p g_{\mu \nu}}  \right) \delta g_{\mu \nu} \, \dd x
\end{align*}
where we used $\delta \dd x=\frac{1}{2} g^{\alpha \beta} \delta g_{\alpha \beta} \dd x$.  By definition a perfect fluid is a continua which admits a   stress-energy tensor of the form
\begin{equation} \label{jus}
T_{\alpha \beta}= \rho\, u_\alpha u_\beta+p (g_{\alpha \beta}+ u_\alpha u_\beta),
\end{equation}
where $u$ is the normalized velocity, $g_{\alpha \beta} u^\alpha u^\beta=-1$, and for barotropic fluids the constitutive relation, $\rho=f(p)$, establishes a functional contraint between the density $\rho$ and the pressure $p$.

\subsection{Barotropic fluids}

Let us show that there is indeed a Lagrangian $L_{(m)}(g_{\alpha \beta} )$ which returns the perfect fluid stress-energy tensor with a chosen-in-advance functional dependence between $\rho$ and $p$. We claim that the action with this property is
\begin{align} \label{act}
S[g_{\alpha \beta}]&=\int \left (\frac{1}{16\pi }(R-2 \Lambda) +  P\left(\sqrt{-g^{\alpha \beta} v_\alpha v_\beta}\,\right) \right) \dd x ,
\end{align}
where function $P(x)$ is determined by the differential equation
\begin{equation} \label{msc}
xP'-P=f(P).
\end{equation}
Observe that the left-had side is $P^*(P'(x))$ if the Legendre transform $P^*$ exists. Equation (\ref{msc}) can be easily integrated since the variables can be separated
\[
x=\exp\left(\int^P \!\!\!\frac{\dd \tilde P}{f(\tilde P)+\tilde P}\right).
\]
At the stationary point the integral curves of $v^\alpha:=g^{\alpha \beta} v_\beta$ are physically interpreted as the flow lines of the fluid, the pressure is $p:=P$ and $\rho=f(p)$.
The variable $x$ is called {\em index of the fluid} and has been proved  very useful in the study of perfect fluids \cite{choquet-bruhat09}, for it is basically the logarithmic acceleration potential (see Euler equation (\ref{hdc}) below). The starting point of the integration is arbitrary, as a consequence $x$ can be redefined up to a factor. It turns out that at the stationary point $v^\alpha=x u^\alpha$, namely $v$ is the {\em dynamical velocity} of the fluid, again a very useful quantity in the study of perfect fluids \cite{choquet-bruhat09}.

In (\ref{act}) $v_\alpha$ is a future directed timelike 1-form field but it {\em is not} a dynamical field with respect to which we need to take a variation (and we stress that the data is the covariant  object $v_\alpha$, not $v^\alpha$). Furthermore, it  could be normalized with respect to $g$ but that does not imply that the variations of $g$ respect the normalization.
 This is the key observation which gives room for an interesting variational principle, for otherwise we would have to replace $g^{\alpha \beta} v_\alpha v_\beta$ with $-1$ in Eq.\ (\ref{act}), obtaining something uninteresting.
 The main idea is that  some terms might be trivial  `on shell'  but variationally non-trivial in general.
 This fact helps to explain why this action passed unnoticed.

Let us prove the claims.
Taking the variation and setting $x=(-g^{\alpha \beta} v_\alpha v_\beta)^{1/2}$
\[
T_{\alpha \beta}=\frac{1}{x} P'(x) \,v_\alpha v_\beta+g_{\alpha \beta} P(x).  \label{jhf}
\]
On the stationary point let us set
\begin{align*}
u_\mu &:= \frac{v_\mu }{ \sqrt{-g^{\alpha \beta} v_\alpha v_\beta} }, \\
\rho&:= x P'(x) -P(x),\\
p&:=P(x),
\end{align*}
where $u^\alpha$ is interpreted as the covariant velocity of the continua and  $\rho$ and $p$ as  density and pressure, respectively. With these definitions $T_{\alpha \beta}$ takes the form (\ref{jus}) and by Eq.\ (\ref{msc}), $\rho=f(p)$ as desired.

\begin{remark}
It is natural to ask if a similar result could be obtained given as data a vector field $v^\alpha$. Indeed, it can be done using as action
\begin{align} \label{ace}
S[g_{\alpha \beta}]&=\int \left (\frac{1}{16\pi }(R-2 \Lambda) +  P\left(\frac{1}{\sqrt{-g_{\alpha \beta} v^\alpha v^\beta}}\right) \right) \dd x ,
\end{align}
where $P(x)$ is related to $f$ as before, at the stationary point $x$ is still the index of the fluid, and $p$ and $\rho$ depend on $x$ as before. However, it is not true that $v^\alpha$ is the dynamical velocity of the fluid, which is why we presented the theory in the 1-form version.
\end{remark}

\subsection{General perfect fluids}
Let us  denote with $n$ the density of baryons, with $T$ the temperature, with $s$ the entropy per baryon (so $v:=1/n$ is the specific volume and $u:=\rho/n$ is the energy per baryon). The first law of thermodynamics is
\cite[Sect.\ 22]{misner73}
\begin{equation} \label{cko}
\dd \left(\frac{\rho}{n}\right)=-p \, \dd \left(\frac{1}{n}\right)+T \dd s
\end{equation}
which can also be rewritten
\[
\dd \rho= \frac{\rho+p}{n}\,\dd n+n T \dd s .
\]
The enthalpy per baryon $h(p,s)$ is the thermodynamic potential defined by $h:=u+p v$, namely
\begin{equation} \label{jud}
h=\frac{\rho+p}{n}.
\end{equation}
From Eq.\ (\ref{cko})
\[
\dd h=\frac{1}{n}\, \dd p+T \dd s .
\]
We can invert $h(p,s)$ so obtaining the function $p=P(h,s)$ which satisfies \cite{schutz70}
\[
\dd p=n \dd h-  nT \dd s ,
\]
thus $p_h)_s=n$, $p_s)_h=-nT$. Using Eq.\ (\ref{jud})  we get $\rho=h p_h-h$.

Given the function $P$ let
\begin{align} \label{acc}
S[g_{\alpha \beta}]&=\int \left (\frac{1}{16\pi }(R-2 \Lambda) +  P\left(\sqrt{-g^{\alpha \beta} v_\alpha v_\beta}\,, s\right) \right) \dd x ,
\end{align}
then by the already presented calculations we obtain that variation with respect to $g_{\alpha \beta}$ gives the stress-energy tensor of the fluid.  It can be observed that the variable $x$ this time is the enthalpy and at the stationary point $v^\alpha = h u^\alpha$ which again is the dynamical velocity (Taub current) in the general case \cite{choquet-bruhat09}. Furthermore, if we consider the equation obtained varying $v_\alpha$ through exact forms (notice that $v_\alpha$ is not necessarily closed), namely $v_\alpha \to v_\alpha+\p_\alpha \varphi$, we get
\[
-\int P_h\,  \frac{g^{\alpha \beta} v_\alpha \p_\beta \varphi}{\sqrt{-g^{\alpha \beta} v_\alpha v_\beta}} \, \dd x=0
\]
which after integration by parts and using $p_h)_s=n$ gives
\begin{equation} \label{mod}
\nabla_\alpha(n u^\alpha)=0,
\end{equation}
which is the conservation of baryons. This approach is essentially that of  \cite{schutz70} (see also \cite{schmid70,brown93}). We shall see in the next section that the Einstein equations imply  $\nabla_u \rho+(\rho+p) \nabla \cdot u=0$. Since the  first principle  built in the function $P$ implies  $\nabla_u \rho=h \nabla_u n+nT \nabla_u s$, we have
\begin{equation} \label{hop}
 h \nabla_\alpha(n u^\alpha) +n T\nabla_u s=0,
\end{equation}
and since we have baryon conservation  we have also entropy conservation along the flow lines. The idea is that entropy cannot increase if we don't have neither heat flow nor  creation of particles. However, in my opinion, it could be incorrect to impose stationarity under variation of $v$. If done one should hope to get the same equations implied by stress-energy conservation or more, not just different ones. In this way we could consider the matter Lagrangian not in pair with the gravitational one.

Schutz goes on to consider a variation of the form $v_\alpha \to v_\alpha+\p_\alpha \varphi+ \theta \p_\alpha s$ where $s$ is the entropy per baryon. The variation with respect to $\varphi$ gives again (\ref{mod}). The variation with respect to $\theta$ gives
\[
\nabla_u s=0
\]
while variation with respect to $s$ gives
\[
\nabla_\alpha (\theta n u^\alpha)-nT=0 \ \Rightarrow \ \nabla_u \theta =T.
\]
The variable $\theta=\int T \dd \tau+ cnst$ is called {\em  thermasy}. Actually, Schutz rather than considering these restricted  variations of $v$ claims that $v$ can be parametrized using potentials of which $s, \varphi,\theta$ are a subset. However, it is strange that  $s$ appears twice, also outside $v$ in the Lagrangian, and furthermore, if the potentials parametrize any $v$ the variation with respect to the potentials should imply the equation obtained through  the variation of $v$ namely $n=0$, which is clearly untenable. I am therefore not entirely convinced that it could be meaningful to vary with respect to the potentials. If one allows for other forms of variations then the option
$v_\alpha \to v_\alpha+\p_\alpha \varphi+ \varphi \frac{1}{h}\p_\alpha s$ is interesting since it gives directly (\ref{hop}).
%
%Variation with respect to $s$ gives
%\[
%0=\nabla_\alpha (\frac{\varphi}{h} n u^\alpha)-nT= -\frac{nT \varphi}{h^2} \nabla_u s+n\nabla_u  (\frac{\varphi}{h})-nT
%\]
%
%We wish to generalize it to the case of heat conduction. One should be careful, however since $\nabla_u \rho+(\rho+p) \nabla \cdot u=0$ does not hold.
%
%The Gibbs free energy per baryon $g(p,T)$ is defined by $g=h-T s$ and satisfies
%\[
%\dd g=\frac{1}{n} \,\dd p-s \dd T.
%\]
%Observe that $\rho=n h-p=ng+nTs-p$.
%The inverse function $p(g,T)$ will be particularly useful. It satisfies
%\[
%\dd p= n \dd g+n s\dd T.
%\]
%The stress-energy tensor in case of heat conduction is
%\[
%T_{\alpha \beta}=\rho u_\alpha u_\beta+p h_{\alpha \beta}+q_\alpha u_\beta+u_\alpha q_\beta
%\]
%which in terms of the entropy vector
%\[
%s_\alpha=ns u_\alpha +\frac{1}{T}\, q_\alpha
%\]
%reads
%\[
%T_{\alpha \beta}=p g_{\alpha \beta}+ (q_\alpha+\frac{1}{2}(\rho+p) u_\alpha) u_\beta+u_\alpha (q_\beta+\frac{1}{2}(\rho+p) u_\beta)
%\]
%or
%\[
%T_{\alpha \beta}=(\rho+p -nsT) u_\alpha u_\beta+p g_{\alpha \beta}+\frac{T}{ns} s_\alpha s_\beta-\frac{1}{nsT}\, q_\alpha q_\beta
%\]
%defined the temperature vector $t_\alpha=\frac{T}{ns} s_\alpha=T u_\alpha+\frac{1}{ns}\, q_\alpha$
%\[
%T_{\alpha \beta}=(\rho+p -nsT) u_\alpha u_\beta+p g_{\alpha \beta}+\frac{ns}{T} t_\alpha t_\beta-\frac{1}{nsT}\,q_\alpha q_\beta
%\]
%The first three terms are obtained from $p(g,T)$ with  the matter Lagrangian
%\[
%P\left(\sqrt{-g^{\alpha \beta} v_\alpha v_\beta}\,, \sqrt{-g^{\alpha \beta} t_\alpha t_\beta}\right)
%\]
% where $v_\alpha =g u_\alpha$. [DA COMPLETARE]

\subsection{The equations of motion}
We have shown that it is possible to obtain the stress-energy tensor of the perfect fluid variationally.
Now, since the left-hand side of the Einstein equation is divergence free (by the naturality of the gravitational action, see \cite[Sect.\ 3.3]{hawking73}), so is the right-hand side, namely
 %Now, from the conservation of the stress-energy tensor, namely from the naturality of the matter action $\int L_{(m)} \dd x$, we get the conservation of energy-momentum,
 $T_{(m) ;\nu}^{\mu \nu}=0$. From here there follow the conservation of mass-energy (continuity equation/first law of thermodynamics)
\begin{equation}
\nabla_u \rho+(\rho+p) \nabla \cdot u=0,
\end{equation}
   and the dynamical equation for the continua \cite[p. 563]{misner73} (Euler's equation)
   \begin{equation} \label{hdc}
   (\rho+p)\, a^\alpha=-h^{\alpha\beta} \nabla_\beta p ,
   \end{equation}
   where $a^\alpha=u^\alpha_{;\beta} u^\beta$ is the acceleration and $h^{\alpha}_{ \beta}=\delta^{\alpha}_{ \beta}+u^\alpha u_\beta$ is the projection on the subspace orthogonal to $u$.
   Since these calculations are well known they will not be repeated here.

\begin{example} $\empty$\\ \indent
For the linear constitutive relation, $f(y)=k y$, we have $P=C x^{1+k}$. Observe that for the vacuum equation of state, namely for $k=-1$, we obtain $P=C$ namely a contribution to the cosmological constant.  For a gas of radiation $k=3$ and $P=Cx^{4}$.
\end{example}

%\begin{example}$\empty$\\ \indent
%For an incompressible fluid of density $\rho_0(=f)$  we have the linear function $P=- \rho_0+ C x$.
%\end{example}

%At this point the reader might have the following concern. If the field $v$ is non-dynamical and fixed since the beginning, does this fact  imply that we are fixing the history of the fluid, and if so what is left of the evolution equations for the dynamics? The answer is that no, we are not fixing the history of the fluid for there are no relative distances between particles with no geometry. Observe that chosen any $p,q\in M$ we can find neighborhoods $U_p$, $U_q$ and a diffeomorphism between them which sends  flow lines of $v$ into flow lines. In other words locally there is no dynamical information in the data provided by the field $v$. The  variational principle introduces the missing local data and allows us  to evolve the spacetime geometry over the fluid so that it becomes possible to speak of fluid motion on spacetime.

We conclude that the dynamical equations for  perfect fluid continua  do admit a simple variational formulation. It is interesting to observe the mechanisms for obtaining these equations requires the Einstein-Hilbert term, namely, the fluid moves as expected but with respect to a spacetime geometry which reacts to the motion of the fluid. Mathematically it could be seen as a drawback since the spacetime geometry is not held fixed, say to the Minkowski form. Physically, however, this could be a satisfactory behavior for the variational formulation seems  admissible precisely under  physically reasonable assumptions. Still one could perhaps fix the geometry in various ways, either introducing constraints or more naturally, taking the limit in which the gravitational constant goes to zero  after the variation, so as to make the influence of matter on geometry negligible.

\section*{Acknowledgments}
I thank Radu Slobodeanu for pointing out several references. \\
%This work has been partially supported by GNFM of INDAM.\\

%\bibliography{../../bibliografie/simultaneity,../../bibliografie/libri,../../bibliografie/miei,../../bibliografie/mieiPreprints,../../bibliografie/mieiProceedings}
%\bibliographystyle{plain}

\end{document}